\documentclass[9pt,twocolumn,twoside]{opticajnl}
\journal{opticajournal} 

\setboolean{shortarticle}{true}

\usepackage{float}
\usepackage{graphicx}
\usepackage{dcolumn}
\usepackage{bm}
\usepackage{hyperref}
\usepackage{siunitx}
\usepackage{setspace}

\renewcommand{\copyright}{\textsuperscript{\tiny{\textcopyright}}}

\title{Optomechanical microgear cavity}
\author[1]{Roberto O. Zurita, Cau\^{e} M. Kersul, Nick J. Schilder, Gustavo S. Wiederhecker}
\author[1,*]{Thiago P. Mayer Alegre}

\affil[1]{Instituto de F\'isica Gleb Wataghin, Universidade Estadual de Campinas (UNICAMP), 13083-859 Campinas, SP, Brazil}
\affil[*]{alegre@unicamp.br}

\begin{abstract}
We introduce a novel optomechanical microgear cavity for both optical and mechanical isotropic materials, featuring a single etch configuration. The design leverages a conjunction of phononic and photonic crystal-like structures to achieve remarkable confinement of both optical and mechanical fields. The microgear cavity we designed in amorphous silicon nitride exhibits a mechanical resonance at \SI{4.8}{\GHz}, and whispering gallery modes in the near-infrared, with scattering-limited quality factors above the reported material limit of up $10^7$. Notably, the optomechanical photoelastic overlap contribution reaches 75\% of the ideal configuration seen in a floating ring structure.
\end{abstract}

\begin{document}
\maketitle

\section{Introduction}

Disk resonators have commonly been used in optomechanics due to their high optical quality factor and ease of integration with photonic integrated circuits via waveguides~\cite{Sun2012, Puckett2021}. In these devices, the high-quality factor (high-Q) optical whispering gallery modes (WGM) couple with the breathing and bending modes of the disk. These mechanical modes oscillate at frequencies in the range of hundreds MHz up to a few GHz for typical microdisk dimensions and have their mechanical strains distributed within the whole disk. However, the strain peak occurs predominantly in the central disk region, away from the WGMs optical fields, leading to rather weak photoelastic interaction. In such structures, the optomechanical interaction is solely given by the moving boundary effect~\cite{DingPRL2010, DingAPL2011, JiangOE2012}. 

To address these limitations, we previously introduced microresonators that use a 1D radial phononic crystal with a bandgap near the desired mechanical frequency~\cite{Santos2017,Carvalho2021}. This approach co-localizes high-Q optical WGM and mechanical modes, increasing the spatial overlap between optical and mechanical modes for stronger optomechanical interaction and reducing the effective motional mass, which enables higher mechanical frequencies, reaching several GHz and supporting the resolved sideband regime. However, this strategy encountered a significant challenge related to the mechanical anisotropy of materials like Si and GaAs, making it difficult to simulate mechanical modes and design a robust phononic bandgap that accommodates changes in crystal orientation along the cavity perimeter. Additionally, the fabrication process for these structures required two distinct dry etch depths: one for the mirror grooves and another to detach the outer disk edge from the material layer, necessitating either a two-step lithography~\cite{Santos2017} process or a study of feature-dependent etch rates~\cite{Carvalho2021}.

In this work, we propose a novel design for an optomechanical microgear cavity based on silicon nitride (Si$_3$N$_4$) requiring a single dry etch step. The optomechanical cavity we propose is a novel ring-like structure that incorporates a two-dimensional (2D) phononic mirror with a compatible microgear-type optical cavity~\cite{Fujita2002,Lu2022}, see Fig.~\ref{fig:device}. This phononic mirror enables high mechanical confinement, while the microgear-inspired tethers ensure effective optical confinement of a whispering gallery-like mode. By combining these two structures, our design achieves simultaneous high optical and mechanical confinement, resembling the behavior of an ideal floating ring, while still being feasible to fabricate. We chose Si$_3$N$_4$ for its well-known optical and mechanical isotropy, which has led to successful demonstrations of optomechanical devices~\cite{Wiederhecker2009, Liu_PRL2013, Grutter2015}. This isotropy has also facilitated recent characterization of one component of its photoelastic tensor~\cite{Gyger2020}, with a measured absolute value of $p_{12}=0.047$. Although this value places Si$_3$N$_4$'s $p_{12}$ between that of silica ($p_{12}=0.270$~\cite{Eggleton2019a}) and silicon ($p_{12}=0.017$~\cite{Eggleton2019a}), it is sufficient to enable a strong contribution of the photoelastic effect to the overall optomechanical interaction.

In the following sections, we will describe the design process for the microgear cavities. We start by discussing the design principles and challenges, followed by an evaluation of our confinement strategies' effectiveness for optical and mechanical modes using Finite Element Method (FEM). Finally, we will compare our design strategy with the ideal floating ring to demonstrate its impact on optomechanical coupling.

\section{Design}

\begin{figure}[t!]
\centering
\includegraphics[width = 8.0cm]{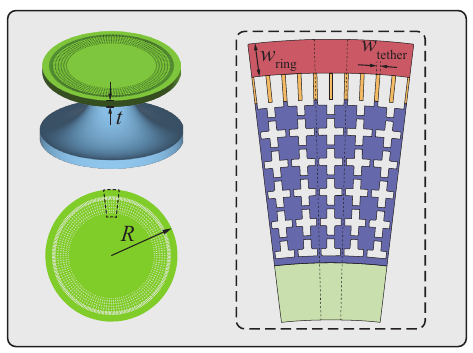}
\caption{\small{Device schematics highlighting the main regions: the outer ring (red), the supporting and floating tethers (orange), phononic mirror (blue). The key dimensions include $R$, representing the disk radius; $t$, the thickness of Si$_3$N$_4$; and $w_\mathrm{ring}$, the width of the ring. The two dashed lines in the right figure highlight the region maintaining rotational symmetry.}}
\label{fig:device}
\end{figure}

Figure~\ref{fig:device} shows the proposed geometry, which can be understood as a ring (highlighted in red) connected to an inner disk (highlighted in blue) by thin tethers (highlighted in orange). The inner disk region, anchored to a pedestal, is patterned with a phononic mirror to confine the mechanical modes within the ring, while the distribution of the supporting tethers is engineered to preserve high-quality factor optical modes.

The tethers supporting the ring act as scattering centers to the optical modes, drastically decreasing their quality factors. Nevertheless, as it is known from the literature on bounded states in the continuum, this effect is strongly mitigated for specific optical modes, when the individual far-field scattering patterns of each scattering center interfere destructively~\cite{Hsu2016}. In the case of a ring-like cavity, it means that the spacing between two neighboring tethers should be such that the driving field has opposite phases between adjacent tethers. For a given optical mode with azimuthal number $m_\mathrm{opt}$ the number of tethers ($N$) should be equal to $2m_\mathrm{opt}$~\cite{Fujita2002,Lu2022}.

Since our design presents an $N-$fold rotational symmetry where the order is given by the number of periods along the azimuthal direction, mechanical and optical modes cannot be properly described by simple 2D axisymmetric simulations, on the other hand, 3D FEM simulations of the entire device would be too computationally demanding. In order to shrink the simulation domain we make use of the cyclic boundary condition for both optical and mechanical fields, allowing us to simulate the full device using only one cyclic unit cell of the structure. In the Supplemental material, we describe the method of creating cyclic boundary conditions for the optical modes. As an example we present a floating ring, where we compare optical and mechanical results obtained using 3D cyclic boundary conditions with those obtained in a regular 2D axisymmetric model.

The phononic mirror is based on a 2D phononic crystal formed by cross-shaped holes~\cite{Kuang2004} known to present a large phononic bandgap. To map this rectangular structure to a disk, we compressed the phononic crystal lateral dimensions as a function of its distance from the disk center ($a_y\rightarrow a_\theta (r) = 2\pi r/N$).
The frequency of the band gap depends essentially on the lattice parameters of the phononic crystal ($a_r$,$a_\theta$), which in the azimuthal direction is set by the number of tethers. One advantage of this approach is that the device can be fabricated using a single etch step, as opposed to previous phononic crystal-assisted cavity designs~\cite{Carvalho2021}.

The outer ring dimensions, $R=\SI{20}{\um}$, $w_\text{ring}=\SI{1}{\um}$ and $t=\SI{0.8}{\um}$, were chosen to be compatible with commercial foundries Si$_3$N$_4$ fabrication processes, that have already been show to produce optical quality factors up to $10^7$~\cite{Pfeiffer2018}. 

In a ring with these dimensions, the largest optomechanical coupling is between the first-order TE mode with the mechanical breathing mode around \SI{4}{\GHz}. To confine this mechanical mode, the mirror needs to have a lattice parameter of around \SI{945}{\nm}, which can be achieved by dividing the ring perimeter into $N=132$ unit cells. As stated before, for this case an improved optical Q-factor can be reached for optical modes with azimuthal order $m_\mathrm{opt}=66$. However, the corresponding optical frequency would be lower than the infrared telecom bands. To ensure that the optical WGM with the highest Q falls within the infrared telecom bands, while hardly affecting the coupling to the phononic mirror, we added one extra floating tether per unit cell, doubling the number of tethers perceived by the optical mode ($N_\text{eff, opt} = 2N$). In this scheme, the first-order TE mode with the highest confinement is the one with $m_\mathrm{opt}=132$ around \SI{190}{\THz}.

\subsection{Optical confinement}

\begin{figure*}[t!]
\centering
\includegraphics[width = 17cm]{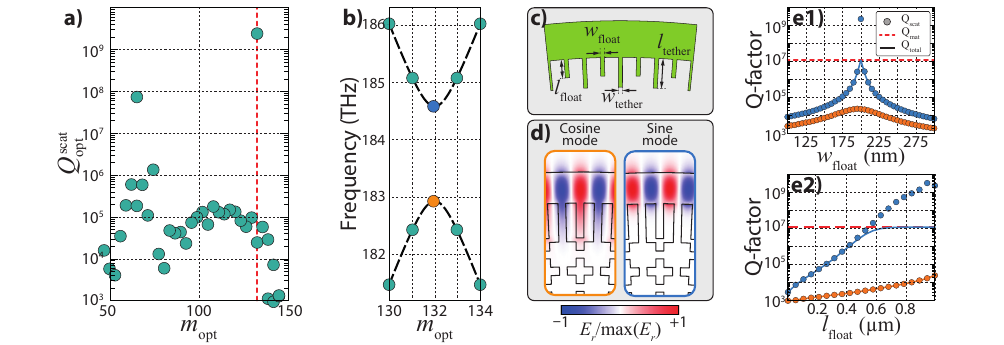}
\caption{\small{  \textbf{(a)} Scattering optical quality factor of the first order TE WGM as a function of its azimuthal number, for a device with $w_\text{tether}=w_\text{float}=$ \SI{200}{\nm} and $l_\text{tether}=l_\text{float}=$ \SI{1}{\um}. A red vertical dashed line is added to highlight the maximum optical quality factor for azimuthal order $m_\text{opt}=132$. \textbf{(b)} Mode splitting of the sine (blue) and cosine (orange) modes. \textbf{(c)} Schematics of the tether region highlighting the most relevant parameters.  \textbf{(d)} Typical mode profile of the Sine (blue) and Cosine (orange) modes in the $r,\theta$ plane. \textbf{(e)} Material and scattering contributions to the total optical quality factor of the ring mode as a function of the floating tether dimensions, where  $Q_\text{total}=(Q_\text{scat}^{-1}+Q_\text{mat}^{-1})^{-1}$ }, for Sine (blue) and Cosine (orange) modes.}
\label{fig:opt_conf}
\end{figure*}

The tethers of the structure introduce scattering that drastically reduces the quality factor of the WGM. However, scattering losses can be reduced if the phase of the optical driving field is opposite between nearest-neighbor tethers~\cite{Hsu2016}. This is achieved in our cavities when the optical mode's azimuthal number equals half the number of tethers, as the scattered light from two adjacent tethers has opposite phases, leading to destructive interference in the far field~\cite{Lu2022}.

Figure~\ref{fig:opt_conf}(a) shows the quality factor of the first radial order TE WGM modes as a function of their azimuthal order. As expected, there is a peak in the quality factor for $m_\text{opt}=N_\text{eff}/2=132$, as highlighted by the vertical red dashed line, where the phase difference between adjacent tethers is $\pi$. Interestingly, another peak is observed at $m_\text{opt}=66$, where the phase difference between adjacent tethers is exactly $\pi/2$. At this point, destructive interference occurs between the scattered fields of a tether and its second nearest neighbor. The Q-factor is lower however, due to an increased misalignment between the tethers, resulting in less effective destructive interference in the far field.

The tethers also form a diffractive grating which couples clockwise (CW) and counter-clockwise (CCW) modes for the optical modes $m_\text{opt}=66$ and  $m_\text{opt}=132$. Figure~\ref{fig:opt_conf}(d) shows the resulting stationary modes, and Fig.~\ref{fig:opt_conf}(b) shows mode splitting of the two resulting stationary modes, which is around \SI{1.6}{\THz}. These modes can be classified according to their electric field distribution within the tether region into sine and cosine modes [Fig. \ref{fig:opt_conf} (d)].  Notably, the higher frequency sine mode exhibits a node within the tether region, resulting in further suppression of the tether-induced scattering, as the electric dipole contribution is zero for this sine mode. For the mode at $m_\text{opt}=132$, this suppression in the sine mode results in a simulated scattering-limited quality factor, $Q_\text{opt}^\text{scat}=4\times10^9$, which is five orders of magnitude larger than the cosine mode.  

Figure~\ref{fig:opt_conf}~(e) highlights the importance of the symmetry between the scattering strengths of the neighboring tethers to $Q_\text{opt}^\text{scat}$. Both length and width mismatch between the tethers can cause drastic changes in $Q_\text{opt}^\text{scat}$ since the different scattering strengths would lead to reduced destructive interference in the far field, therefore resulting in increased scattering losses. 

We define the material and fabrication defects contributions to the optical Q-factor as $Q_\text{mat}$, which typical value is indicated by a red horizontal dashed line in Fig.~\ref{fig:opt_conf}~(e1). Using this as an upper bound for our quality factor leads us to evaluate the resilience of our design to the tether parameters. As expected, the sine mode is more resilient to a length mismatch between the tethers than the cosine mode, as the improved confinement makes it less prone to be affected by length mismatch that occurs on the side of the tether away from the ring. However, both modes are sensitive to a width mismatch, as it affects the electromagnetic environment directly next to the ring, where a significant part of the optical mode is located. This means that for the device to be no longer limited by the scattering introduced by the design, the tolerance for the width mismatch needs to be under \SI{10}{\nm}.

\subsection{Mechanical confinement}

\begin{figure*}[ht!]
\centering
\includegraphics[width = 17.0cm]{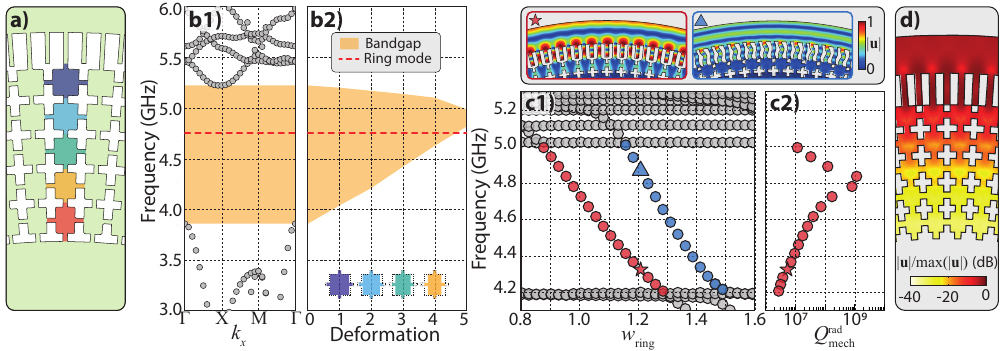}
\caption{\small{\textbf{(a)} Device schematics highlight of the mechanical crystal-like mirror and the difference between each of its radial periods. \textbf{(b)} Band diagram for the 2D square lattice phononic mirror ($a=\SI{830}{\nm}$) as well as the evolution of its bandgap as we deform the horizontal axis to match the mean azimuthal lattice parameter of each radial period in figure \textbf{(a)}. \textbf{(c1)} Mechanical frequency of modes as a function of the width of the ring. The top insets show the breathing and azimuthal modes' displacement profiles highlighted in red (star) and blue (triangle) respectively. \textbf{(c2)} Quality factor of the breathing mode for each geometry in \textbf{(c1)} . \textbf{(d)} Mechanical mode profile in log scale for the breathing mode highlighted in red  (star) in figure \textbf{(c)} normalized by its maximum value.}}
\label{fig:mech_conf}
\end{figure*}

To design the phononic mirror, we begin by designing a square phononic crystal in the $xy$-plane that has a bandgap around the breathing mode frequency and then conformally map it to fit inside a circular shape [Fig.~\ref{fig:mech_conf} (a)]. When changing the square structure to a $r,\theta$-space, the most affected parameter is the lattice parameter ($a$). In the original crystal $a_x=a_y=a$, however, in order to fit the structure in the circle, the lattice parameter in the azimuthal direction ($a_\theta$) changes with the radius. This change is given by $N a_\theta = 2\pi r$. Since $N$ is fixed, we design the conformed structure based on $a_r$ and $N$ as our design parameters.

To conformally map the crystal to the $r\theta$-space, we first calculate $N$ as a function of the square lattice parameter ($a_0$). Since $N$ needs to be integer, we chose the one closest to $2\pi r_0/a_0$, where $r_0$ is the radius of the phononic crystal region ($R-w_\text{ring}$). The radial lattice parameter is then calculated as $a_r = 2\pi r_0/N$ so that the crystal's outmost period resembles the ideal square lattice.

In the resulting structure, every azimuthal neighbor is identical, but every inner cell is a bit deformed compared to its outer neighbor. In order to investigate how this deformation affects the efficiency of the phononic mirror, we simulate how the bandgap would evolve with each cell by calculating a rectangular phononic crystal with $a_y=a_r$ and $a_x=\text{mean}\{a_\theta\}$ for each radial period (Fig.~\ref{fig:mech_conf} (b)). In the graph, we can see that the deformation shrinks the bandgap. Beyond the fifth period, the deformation is so large that the expected mechanical frequency falls outside of this cells' bandgap, meaning that after five periods the mechanical confinement would stop improving.

Although this structure cannot be considered a proper 2D phononic crystal due to its lack of discrete translational symmetry in the radial direction, we expect its functioning to confine mechanical modes in the outer ring to remain. This is evidenced by the simulations of the device's mechanical modes shown in Fig.~\ref{fig:mech_conf} (c). In these simulations, we calculated the mechanical modes for the entire structure as a function of the ring's width. Modes from the shield are not affected by this change, so the lack of modes with constant frequency indicates a bandgap around \SI{4.6}{\GHz}.

As the width of the ring increases, the frequency of the breathing mode (highlighted in red) decreases and eventually enters the bandgap at approximately \SI{0.9}{\um}, exiting it around \SI{1.3}{\um}. When the mode enters the bandgap, its radiation-limited Q-factor ($Q_\text{mech}^\text{rad}$) rapidly increases to $10^9$ at \SI{4.8}{\GHz}, and then gradually decreases to $10^6$ at \SI{4.2}{\GHz}. The frequency dependence of the bandgap's effectiveness is a direct result of the phononic mirror having a limited number of periods. However, by placing the mechanical mode close to the center of the bandgap, we can ensure a $Q_\text{mech}^\text{rad}>10^8$ (Fig.~\ref{fig:mech_conf}(d)) with a high resilience to fabrication mismatch regarding the ring width. 

\subsection{Optomechanical coupling}

The proposed optomechanical microgear cavity successfully confines optical and mechanical waves within the outer ring (Figs.~\ref{fig:opt_conf} and \ref{fig:mech_conf}). However, the inclusion of tethers, both anchored and floating, inevitably alters the profile of the mechanical mode. Specifically, for the breathing mode, a significant portion of the mechanical displacement is concentrated in the floating tether (Fig.~\ref{fig:mech_conf}(d)). Therefore, it is crucial to investigate how this perturbation in the mode profile affects the cavity's optomechanical response.

The coupling between the optical and mechanical modes is characterized by the optomechanical coupling factor ($g_0$), which typically consists of the sum of the moving boundary (MB) and photoelastic (PE) contributions: $g_0 = g_\mathrm{MB} + g_\mathrm{PE}$, where $g_\mathrm{MB}$ and $g_\mathrm{PE}$ are determined by spatial overlap integrals between the optical and mechanical modes for the moving boundary and photoelastic effects, respectively~\cite{Wiederhecker2019c}.

The photoelastic effect is determined by the photoelastic tensor ($\mathbf{p}$), which quantifies the change in optical permittivity resulting from applied strain in the material~\cite{Boyd2003}. In isotropic materials like amorphous silicon nitride, the photoelastic tensor $\mathbf{p}$ has only two independent components which in the Voigt notation are $p_{11}$ and $p_{12}$~\cite{auld1990acoustic}. While the absolute value of $p_{12}$ in thin films of $\text{Si}_3\text{N}_4$ has been measured in a recent study on Brillouin scattering \cite{Gyger2020}, the value of $p_{11}$ remains unknown. To study the optomechanical interaction within our cavities without complete knowledge of the two components of $\mathbf{p}$, we introduce the coupling factors $g_{11}$ and $g_{12}$, such that $g_\text{PE} = p_{11}g_{11}+p_{12}g_{12}$

\begin{table*}[h]
\centering
\caption{Comparison between the floating ring and microgear performance for $w_\text{tether}=$\SI{200}{\nm}. Notice that in practice the total optical$(*)$ and mechanical quality$(**)$ factors will be limited by material and thermal effects, respectively.}
\label{tab:comp}
\begin{tabular}{|c|c|c|c|c|c|}
\hline
\textbf{Device}           & \textbf{$Q_\text{opt}^\text{scat {*}}$}       & \textbf{$Q_\text{mech}^\text{rad {**}}$}      &\textbf{$g_\text{MB}/(2\pi)~\text{(kHz)}$} & \textbf{$g_\text{11}/(2\pi)~\text{(kHz)}$} & \textbf{$g_\text{12}/(2\pi)~\text{(kHz)}$} \\ \hline
\hline
Floating ring       & $6.2\times10^{15}$ & -                   & 2.40          & -79.6             & 48.5             \\ \hline
Microgear (Sine)    & $1.9\times10^{9}$  & $6.5\times10^{8}$ & 0.80           & -59.9             & 39.9             \\ \hline
Microgear (Cosine) & $1.9\times10^{4}$  & $6.5\times10^{8}$ & 0.45           & -48.6             & 32.4             \\ \hline
\end{tabular}
\end{table*}

Table \ref{tab:comp} presents a performance comparison between the optomechanical microgear and the ideal floating ring. The optomechanical coupling between the mechanical breathing mode and the optical cosine mode is slightly lower than between the mechanical breathing mode and the optical sine mode, which can be attributed to the larger impact of the tethers for the overlap between the mechanical and the cosine modes.

When comparing the moving boundary and photoelastic coupling of the sine (cosine) mode in the microgear with the optimal case of the floating ring, we find that $g_\text{MB}$ for the microgear is 35\%(19\%) of the ideal case, while $g_\text{PE}$ is approximately 75\%(61\%). As expected, the addition of tethers has a greater impact on the moving boundary coupling, particularly on the cosine mode, due to the reduction of the electric field strength at the geometry's edges, resulting in reduced radiation pressure. The photoelastic coupling is less affected since the change in both the optical and mechanical modes is less pronounced within the inside of the ring.

\begin{figure}[ht!]
    \centering
    \includegraphics[width = 8.0cm]{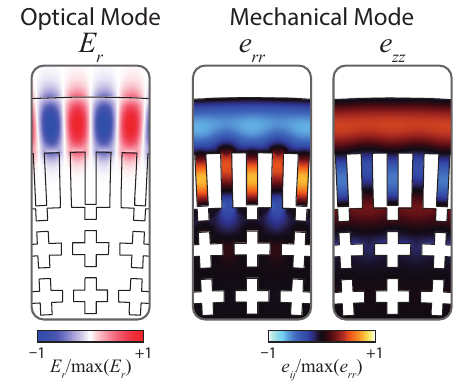}
    \caption{Mode profile for the optimized optical sine mode and mechanical breathing mode. The strain fields are both normalized by the max($e_{rr}$).}
    \label{fig:strain}
\end{figure}

The optomechanical coupling for $g_{11}$ and $g_{12}$ is relatively similar due to the profile and magnitude similarity of the strain fields of the breathing mode in the $rr$ and $zz$ directions, as shown in Fig.~\ref{fig:strain}. This allows the dominant electric field component, $E_r$, to couple with $e_{rr}$ via $p_{11}$ and with $e_{zz}$ via $p_{12}$ with similar strengths. The strain fields $e_{rr}$ and $e_{zz}$ have opposite sign due to the Poisson effect. This means that the relative sign between $p_{11}$ and $p_{12}$ is important in evaluating the value of $g_\text{PE}$. In most of the materials used in optomechanics like silica, GaAs, and $\text{As}_2\text{S}_3$, $p_{11}$ and $p_{12}$ have equal sign \cite{Eggleton2019a,Lide2004}. However, in materials like silicon, the photoelastic constants have opposite signs, which in this case would enhance $g_\text{PE}$. 

To consider different scenarios, we examined three cases: $|p_{11}|\ll |p_{12}|$, $p_{11}=p_{12}$, and $p_{11}=-p_{12}$. In these cases, $g_\text{PE}$ is \SI{1.88}{\kHz}, \SI{0.94}{\kHz}, and \SI{4.70}{\kHz}, respectively. In all three scenarios $g_\text{PE}$ is comparable to $g_\text{MB}$ and, therefore, this cavity can be useful in studying photoelasticity in thin layers of amorphous silicon nitride.

\section{Conclusion}

In summary, we have designed a Si$_3$N$_4$ cavity for optomechanics that achieves confinement of both optical and mechanical waves within a compact region. The fabrication of this cavity requires a single etching step and its design parameters are compatible with currently available foundry technology. For the mechanical confinement of the \SI{}{\GHz} breathing mode, we used a circular phononic-crystal like structure, which successfully achieved confinement even without a proper radial translational symmetry. On the other hand, optical confinement was tailored to enable a high-Q whispering gallery mode with $m_\mathrm{opt}=132$ at \SI{190}{\THz}. In both cases, the limitations on the quality factor are primarily imposed by material properties and fabrication defects rather than design-related mechanisms such as scattering losses.

The designed cavity approaches the ideal case of a floating ring with comparable losses (limited by material and fabrication) and achieves optomechanical coupling that is 70\% below the ideal case for $g_\text{MB}$ and 25\% for $g_\text{PE}$. To mitigate the coupling mismatch, further reduction in $w_\text{tether}$ can be pursued through improvements in fabrication limits or by employing a wet etch with acid, a process already required for structure release.

The significant contribution of $g_\text{PE}$ in the optomechanical coupling within these structures provides an opportunity to explore the photoelastic effect in Si$_3$N$_4$. Measuring the coupling between TE and TM whispering gallery modes and the breathing mode can provide valuable information about the magnitude and relative sign of the components. This exploration is particularly intriguing since the structure also allows for high-quality TM modes with the same azimuthal symmetry as TE modes. The photoelastic coupling in TM modes should exhibit similarities to TE modes, with the only difference being that $g_{11}$ and $g_{12}$ interchange due to the polarization change of the optical mode.

Finally, the addition of the tether introduces a unique feature not commonly observed in single-disk cavity optomechanics—tailorable optical coupling between two modes: the sine and cosine modes. The separation between these modes, such as the \SI{1.6}{\THz} case considered here, can be adjusted by engineering the scattering strength, achieved by modifying the tether or ring width~\cite{Lu2022}. This characteristic opens up possibilities for applications in the realm of multimodal optomechanics phenomena such as cavity-enhanced sideband coupling and exceptional point optomechanics.

\section*{Funding} 
This work was supported by S\~{a}o Paulo Research Foundation (FAPESP) through grants 
19/13564-6, 
20/06348-2, 
22/06254-3, 
18/15580-6, 
18/15577-5, 
18/25339-4, 
Coordena\c{c}\~{a}o de Aperfei\c{c}oamento de Pessoal de N\'{i}vel Superior - Brasil (CAPES) (Finance Code 001), and Financiadora de Estudos e Projetos (Finep).

\section*{Disclosures} 
The authors declare no conflicts of interest.

\section*{Data availability}
Data underlying the results presented in this paper are available at ZENODO$^\text{\textregistered}$ repository (\href{https://doi.org/10.5281/zenodo.7363097}{10.5281/zenodo.7363097})~\cite{Zenodo_MicroGear}, including FEM simulations, scripts files for generating figures and data related to the optimization method.

\bibliography{bib_microgear}

\end{document}


\title{Supplementary Material: Optomechanical microgear cavity}

\author{Roberto O. Zurita~\authormark{1}, Cau\^{e} M. Kersul~\authormark{1}, Nick J. Schilder~\authormark{1} Gustavo S. Wiederhecker~\authormark{1}, and Thiago P. Mayer Alegre~\authormark{1,*}}

\address{
\authormark{1}Instituto de F\'isica Gleb Wataghin, Universidade Estadual de Campinas (UNICAMP), 13083-859 Campinas, SP, Brazil\\
}

\email{\authormark{*}alegre@unicamp.br} 


\section*{Eletromagnetic cyclic boundary condition}

In order to simulate the optical and mechanical fields of the microgear cavity, we have used finite element method (FEM) simulations. In particular, we use the discrete azimuthal symmetry of our design to shrink the simulation domain to a single azimuthal unit cell between two attached tethers. As we know, the eigenmodes must have the same symmetries as the system itself, in such a way that a rotation by an angle $\theta = 2\pi/N$, where $N$ is the number of unit cells, only multiplies the fields by a complex phase. This means that for any eigenmode of the system, we can relate the values of a vector field, $\vec{F}$, at the edges of the unit cell using:

\begin{equation}
\vec{F}_L = \bm R (\theta) \vec{F}_R e^{i m\theta},
\label{eq1}
\end{equation}
where $\vec{F}_{L/R}$ represents the fields at the left/right edges of the unit cell, $\bm R (\theta)$ is the rotation matrix representing an anti-clockwise rotation by an angle $\theta$ with respect to the $z-$axis, and $m$ is the eigen-mode azimuthal order. \textit{À priori} $m$ could be any real number, nevertheless as the resonator is closed we must have:

\begin{equation}
\vec{F}_R = \bm R (2\pi) \vec{F}_R (e^{i m\theta})^N.
\end{equation}
As $\bm R (2\pi) = \bm I$, it implies that $e^{i m N \theta} = 1$, and, consequentially:

\begin{equation}
m N \theta = 2\pi l\textrm{,   with   }  l \in \mathbb{Z}.
\end{equation}
Finally, replacing $\theta = 2\pi/N$ we arrive at:

\begin{equation}
m  \in \mathbb{Z}.
\end{equation}
In other words, as our system is finite, its eigen-modes are classified through a discrete set of azimuthal numbers.

We can also show that an arbitrary azimuthal order $m_i$ can always be reduced to an azimuthal order $m$ such that $N/2 > m > -N/2$. Writing the unit cell phase factor in this case:

\begin{equation}
e^{i m_i \theta} = e^{i (m + bN) \theta} \textrm{,   with   }  b \in \mathbb{Z}.
\end{equation}
As $b$ can be any integer, we can always choose it in such a way that $N/2 > m > -N/2$, in such a way that $e^{i b N \theta} = 1$, and:

\begin{equation}
e^{i m_i \theta} = e^{i m \theta}.
\end{equation}
This allows us to define the first Brillouin zone by  $-N/2 \geq m \geq N/2$. 

For the optical modes, the vector field used was the electric field tangential to the boundaries of the unit cell:
\begin{equation}
\vec{F}_{\textrm{opt}} = \vec{E}-(\vec{E}\cdot\hat{n})\hat{n},
\end{equation}
where $\hat{n}$ is the surface normal. 

For the mechanical modes, it was the mechanical displacement:

\begin{equation}
    \vec{F}_{\textrm{mech}} = \vec{u}.
\end{equation}

The condition given by Eq.\ref{eq1} is implemented in the FEM simulation through the addition of a boundary constraint in the form of a Dirichlet boundary condition involving the fields at the two boundaries of the unit cell.

\section*{Floating ring simulation: Cyclic boundary condition vs axis-symmetric simulations}

\begin{figure}[ht!]
\centering
\includegraphics[width = 13.0cm]{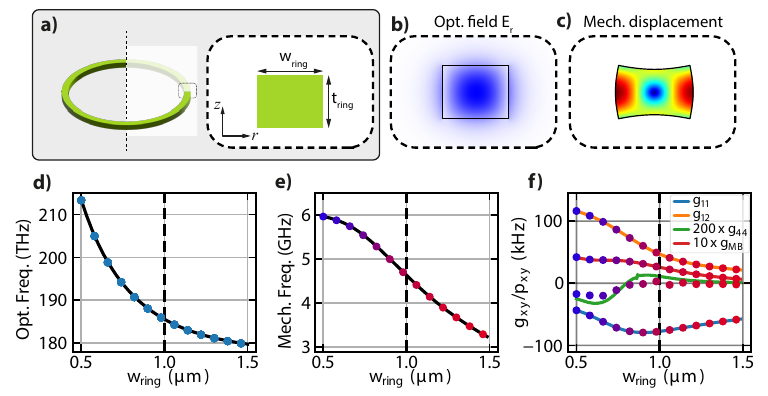}
\vspace{-10pt}
\caption{\small\textbf{a} Floating ring with its cross-section. \textbf{b} Optical mode electric field in the $r-$direction of a fundamental TE mode. \textbf{c} Mechanical mode displacement profile of the fundamental breathing mode. \textbf{d}, \textbf{e} and \textbf{f} present, respectively, how the fundamental TE mode frequency, the fundamental breathing mode frequency, and the different optomechanical coupling components, vary with the ring width. The lines present the results of the 2D axis-symmetric simulations, while the dots are results from the 3D simulations using the cyclic boundary condition. In \textbf{e} and \textbf{f}, the color of the dots indicates the ratio between the vertical (blue) and horizontal (red) components of the stress tensor. In all the simulations, the ring height, the ring radius, and the optical mode azimuthal order are set to be, respectively, $\SI{800}{nm}$, $\SI{20}{\micro\meter}$ and $132$, the same as in the main text.}
\label{fig:supp1}
\end{figure}

As a way to verify the implementation of the cyclic boundary condition in 3D, we have compared its results to well-established axis-symmetric 2D simulations for the floating ring, as presented in Fig.~\ref{fig:supp1}.

In Fig.~\ref{fig:supp1} (a), we show the floating ring and its cross-section. In Fig.~\ref{fig:supp1} (b), we show the radial electric field profile of the first-order TE optical mode, while in Fig.~\ref{fig:supp1} (c), we do the same for the mechanical displacement of the first-order breathing mode. In Figs.~\ref{fig:supp1} (d) and (e), we present how the optical and mechanical modes of interest vary as a function of the ring width, where we see that our 3D simulations using the cyclic boundary condition (dots) agree very well with the 2D axis-symmetric simulation (solid lines). As the ring width increases, the optical mode gets more confined within the dielectric ring, leading to a higher effective refractive index, and consequently, a smaller optical frequency. The behavior of the mechanical mode is more complex, for rings, with $\textrm{w}_{\textrm{ring}}$ smaller than \SI{1}{\micro\meter} the fundamental mode is dominated by vertical contributions, as the width of the ring increases, the frequency of the mechanical mode decreases and its horizontal component increases. For $\textrm{w}_{\textrm{ring}}$ larger than \SI{1}{\micro\meter} the horizontal component is predominant.

In Fig.~\ref{fig:supp1} (f), we present how the different components of the optomechanical coupling vary with the ring width. The $g_{12}$ component is dominant for thin rings, as in this case, the dominant optical field component is $E_r$, while the dominant mechanical field component is $e_{zz}$. As $\textrm{w}_{\textrm{ring}}$ increases to $\SI{1}{\micro\meter}$, $S_{rr}$ becomes dominant, and consequentially the $g_{11}$ component increases while $g_{12}$ decreases. For $\textrm{w}_{\textrm{ring}}$ larger than $\SI{1}{\micro\meter}$, the overall optical and mechanical mode volumes increase in such a way that all components of the optomechanical coupling decrease, such effect is particularly important for $g_\text{MB}$. The $g_{44}$ is at least 500 times smaller than the $g_{11}$ and $g_{12}$ components, as the shear components of the breathing mechanical mode are very small. Again 3D simulations agree very well with the 2D simulations, with a small exception for $g_{44}$, where the calculated values are within the numerical error of our simulation.

Data underlying the results presented in this paper are available at ZENODO$^\text{\textregistered}$ repository (\href{https://doi.org/10.5281/zenodo.7363097}{10.5281/zenodo.7363097})~\cite{Zenodo_MicroGear}, including FEM simulations, scripts files for generating figures and data related to the optimization method.

\bibliography{bib_microgear}